# Adiabatic Quantum Computing with Phase Modulated Laser Pulses

Debabrata Goswami

Indian Institute of Technology, Kanpur – 208016, INDIA

Email: dgoswami@iitk.ac.in

Implementation of quantum logical gates for multilevel system is demonstrated through decoherence control under the quantum adiabatic method using simple phase modulated laser pulses. We make use of selective population inversion and Hamiltonian evolution with time to achieve such goals robustly instead of the standard unitary transformation language.

Use of adiabatic evolution for quantum computation has recently become an attractive approach due to its inherent robustness[1,2,3,4]. In the framework of adiabatic quantum method, logical implementation of quantum gates uses the language of ground states, spectral gaps and Hamiltonians wherein a quantum gate represents a device which performs a unitary transformation on selected qubits in a fixed period of time. Thus, a computational procedure in the adiabatic quantum computation model is described by the continuous time evolution of a time-dependent Hamiltonian with limited energetic resources—an aspect that is often neglected in the unitary gate language[5].

In this paper, we show that an important aspect of the adiabatic quantum computation model lies in addressing an atomic or molecular ensemble and hence in robust implementation. We first demonstrate a simple Hadamard operation with phase modulated laser pulses. Next we show how selective population transfer in a three-level system that has also been demonstrated experimentally[6,7] can be a very useful adiabatic quantum computing logic. Finally, we show that it is possible to decouple states that are parts of the coupled vibrational relaxation tier into simple qubits through control of decoherence through adiabatic coupling. As far as we know, these results are the first realistic demonstration of the possibility of using ensemble states for adiabatic quantum computation in multilevel systems.

We apply a linearly polarized laser pulse of the form $E(t) = \mathcal{E}(t)e^{i[\omega t + \phi(t)]}$ to a simple two-level system with |1>→|2> transition, where |1> and |2> represent the ground and excited eigenlevels, respectively, of the field-free Hamiltonian. The laser carrier frequency or the center frequency for pulsed lasers is $\omega$. We have $\varepsilon(t)$ and $\phi(t)$ as the instantaneous amplitude and phase. We can define the rate of change of instantaneous phase, $\dot{\phi}(t)$ as the frequency-sweep. If we expand the instantaneous phase function of $E(t)$ as a Taylor series with constants $b_n$, we have

$$\phi(t) = b_0 + b_1 t + b_2 t^2 + b_3 t^3 + b_4 t^4 + b_5 t^5 + \ldots.$$

$$\dot{\phi}(t) = \quad b_1 + 2b_2 t + 3b_3 t^2 + 4b_4 t^3 + 5b_5 t^4 + \ldots. \qquad (1)$$

$$\dot{\phi}(t) = \sum_{n=1} n b_n t^{(n-1)}$$

In a recent paper[8], we have proposed the use of simple chirped pulses, which, by contrast, have been produced routinely at very high intensities and at various different wavelengths for many applications, including selective excitation of molecules in coherent control. Establishing this generalization enables us to treat all possible chirped pulse cases by exploring the effects of each of the terms in Eqn. (1) initially for a simple two-level system and then extend it to the multilevel situation for a model five-level system of anthracene molecule, which has been previously investigated with complicated shaped-pulses[9,10]. We use a density matrix approach by numerically integrating the Liouville equation $\dfrac{d\rho(t)}{dt} = \dfrac{i}{\hbar}[\rho(t), H^{FM}(t)]$ for a Hamiltonian in the rotating Frequency Modulated (FM) frame of reference. $\rho(t)$ is a 2×2 density matrix whose diagonal elements represent populations in the ground and excited states and off-diagonal elements represent coherent superposition of states. The Hamiltonian for the simple case of a two-level system under the effect of an applied laser field can be written in the FM frame for N-photon transition[11] as, $H^{FM} = \hbar \begin{pmatrix} \Delta + N\dot{\phi}(t) & \dfrac{\Omega_1}{2} \\ \dfrac{\Omega_1^*}{2} & 0 \end{pmatrix}$. The time derivative of the phase function, $\dot{\phi}(t)$, appears as an additional resonance offset over and above the time-independent detuning $\Delta = \omega_R - N\omega$, while the direction of the field in the orthogonal plane remains fixed. We define the multiphoton Rabi Frequencies as complex conjugate pairs: $\Omega_1(t) = k(\mu_{eff}.\varepsilon(t))^N/\hbar$ and $\Omega_1^*(t) = k(\mu_{eff}.\varepsilon^*(t))^N/\hbar$, where $k$ is a proportionality constant having dimensions of (energy)$^{(1-N)}$, which in SI units would be Joule$^{(1-N)}$. For the $|1\rangle \rightarrow |2\rangle$ transition, $\omega_R = \omega_2 - \omega_1$ is the single-photon resonance frequency. We have assumed that the transient dipole moment of the individual intermediate virtual states in the multiphoton ladder result in an effective transition dipole moment, $\mu_{eff}^N$, which is a product of the individual N virtual state dipole moments, $\mu_N$, (i.e., $\mu_{eff}^N = \prod_n^N \mu_n$). This approximation is particularly valid when intermediate virtual level dynamics for multiphoton interaction can be neglected[3,12].

Let us extend the two-level formalism first to a three-level system of alkali atom excitations. The Hamiltonian for such a simple case of a three-level system under the effect of an applied laser field can be written in the FM frame for N-photon transition as,

$$\begin{array}{ccc} |0\rangle & |1\rangle & |2\rangle \end{array}$$
$$H^{FM} = \hbar \begin{pmatrix} 0 & \Omega_{01}(t) & \Omega_{02}(t) \\ \Omega_{01}^*(t) & \Delta_1 + N\dot{\phi}(t) & 0 \\ \Omega_{02}^*(t) & 0 & \Delta_2 + N\dot{\phi}(t) \end{pmatrix}, \text{ where, } \Omega_{01}(t) \text{ is the transition matrix element}$$

between the ground state |0> and the excited state |1> while $\Omega_{02}(t)$ is the transition matrix element between the ground state |0> and the excited state |2>, expressed in Rabi frequency units. In such a model with two possible transitions (Fig.1a), femtosecond pulses have enough bandwidth to excite both the transitions and there is hardly any selectivity possible. At the end of the pulse, population gets equally distributed between the three states |0>, |1> and |2>, if the coupling for |0> → |1> and |0> → |2> are identical (Fig.1b). However, as show in Fig. 1c, in case of an adiabatic population transfer process between the coupled states, it is possible to selectively excite either state |1> or state |2> by simple linear frequency sweeping the laser frequency either

from red to blue or from blue to red. This has been demonstrated experimentally also in case of sodium and rubidium atomic transitions[6,7].

Finally, we extend the formalism to a multilevel situation involving intramolecular vibrational relaxation (IVR). In the conventional zeroth order description of intramolecular dynamics, the system can be factored into an excited state that is radiatively coupled to the ground state, and nonradiatively to other bath states that are optically inactive (Fig. 2a). These "dark" states have no radiative transition moment from the ground state as determined by optical selection rules[13]. They can belong to very different vibrational modes in the same electronic state as the "bright" state, or can belong to different electronic manifolds. These dark states can be coupled to the bright state through anharmonic or vibronic couplings. Energy flows through these couplings and the apparent bright state population disappears. Equivalently, the oscillator strength is distributed among many eigenstates. The general multilevel Hamiltonian in the FM frame for an N-photon transition (N≥1), expressed in the zero-order basis set, is:

$$H = \hbar \begin{pmatrix} & |0\rangle & |1\rangle & |2\rangle & |3\rangle & |4\rangle & \cdots \\ & 0 & \Omega_1(t) & 0 & 0 & 0 & \cdots \\ & \Omega_1^*(t) & \delta_1(t) & V_{12} & V_{13} & V_{14} & \cdots \\ & 0 & V_{12} & \delta_2(t) & V_{23} & V_{24} & \cdots \\ & 0 & V_{13} & V_{23} & \delta_3(t) & V_{34} & \cdots \\ & 0 & V_{14} & V_{24} & V_{34} & \delta_4(t) & \cdots \\ & \vdots & \vdots & \vdots & \vdots & \vdots & \end{pmatrix} \qquad (2)$$

where, $\Omega_1(t)$ (and its complex conjugate pair, $\Omega_1^*(t)$) is the transition matrix element expressed in Rabi frequency units, between the ground state $|0\rangle$ and the excited state $|1\rangle$. The background levels $|2\rangle$, $|3\rangle$,... are coupled to $|1\rangle$ through the matrix elements $V_{12}$, $V_{23}$, etc. Both Rabi frequency $\Omega_1(t)$ and detuning frequency $[\delta_{1,2,...} = \Delta_{1,2,...} + N\dot{\phi}(t)]$ are time dependent (the time dependence is completely controlled by the experimenter). In general, the applied field would couple some of the dark states together, or would couple $|1\rangle$ to dark states, and thus, the $V_{ij}$ terms would have both an intramolecular, time independent component and a field-dependent component. As an alternative to Eqn. (2), the excited states' submatrix containing the bright state $|1\rangle$ and the bath states $|2\rangle$, $|3\rangle$,... can be diagonalized to give the eigenstate representation containing a set of $\Delta_i'$ as diagonal elements and corresponding $\Omega_i'$ as off-diagonal elements. The eigenvalues of such a time-dependent Hamiltonian representation is often referred to as the dressed states of the system. Such a representation corresponds closely to what is observed in conventional absorption spectroscopy. As long as the intensity of the field is very low ($|\Omega_i'| << \Delta_i'$) the oscillator strength from the ground state (and hence the intensity of the transition, which is proportional to $|\Omega_i'|^2$ is distributed over the eigenstates, and the spectrum mirrors the distribution of the dipole moment. On the other hand, a pulsed excitation creates a coherent superposition of the eigenstates within the pulse bandwidth. Physically, in fact, the presence of the dark states has been key to the loss of selectivity of excitation to a specified bright state. Interestingly such a process essentially is a Hadamard operation in Quantum Computing Language as this enables us to produce equal superposition between the ground and excited states which form the qubits.

Another common situation with short pulses is a ladder excitation situation where the individual excited states undergo dephasing through a coupled energy structure with states $|0\rangle$, $|1\rangle$, $|2\rangle$, etc. in the zero-order basis as shown in Fig. 2b. Such a model of IVR is often referred to as the tier model and is common in polyatomic molecules and in most rovibrational states[14] and can be represented by the following Hamiltonian:

$$
H^{FM} = \hbar
\begin{array}{c}
\begin{array}{cccccccccc}
|0\rangle & |1\rangle & |2\rangle & |3\rangle & |4\rangle & |5\rangle & |6\rangle & |7\rangle & |8\rangle & |9\rangle
\end{array} \\
\begin{pmatrix}
0 & \Omega_1(t) & \Omega_2(t) & \Omega_3(t) & 0 & 0 & 0 & 0 & 0 & 0 \\
\Omega_1^*(t) & \delta_1(t) & V_{12} & V_{13} & V_{14} & V_{15} & 0 & 0 & 0 & 0 \\
\Omega_2^*(t) & V_{12} & \delta_2(t) & V_{23} & V_{24} & V_{25} & V_{26} & V_{27} & 0 & 0 \\
\Omega_3^*(t) & V_{13} & V_{23} & \delta_3(t) & 0 & 0 & V_{36} & V_{37} & V_{38} & V_{39} \\
0 & V_{14} & V_{24} & 0 & \delta_4(t) & 0 & 0 & 0 & 0 & 0 \\
0 & V_{15} & V_{25} & 0 & 0 & \delta_5(t) & 0 & 0 & 0 & 0 \\
0 & 0 & V_{26} & V_{36} & 0 & 0 & \delta_6(t) & 0 & 0 & 0 \\
0 & 0 & V_{27} & V_{37} & 0 & 0 & 0 & \delta_7(t) & 0 & 0 \\
0 & 0 & 0 & V_{38} & 0 & 0 & 0 & 0 & \delta_8(t) & 0 \\
0 & 0 & 0 & V_{39} & 0 & 0 & 0 & 0 & 0 & \delta_9(t)
\end{pmatrix}
\end{array}
\qquad (3)
$$

A short pulse laser can optically couple the states |1>, |2>, |3> etc. to the ground state |0> with respective transition matrix elements expressed in Rabi units as $\Omega_1(t)$, $\Omega_2(t)$, $\Omega_3(t)$, etc. and their corresponding conjugates. The background levels |4>, |5>, |6>, etc. are coupled to the optically excited states |1>, |2> and |3> through the matrix elements $V_{14}$, $V_{15}$, $V_{24}$, etc. Such a molecular system can become useful for realizing qubits[5] effectively if these large number of optically coupled states can be accessed simultaneously as has been in case of the atomic system of Rydburg state of cesium[15]. However, the difficulty in extending this scheme to the molecular system starts at the very first step of initializing the qubits due to high decoherence of the possible qubit states as in the gdanken system Hamiltonian presented in Eqn. (3). Thus an adiabatic scheme is necessary.

From experimental results on the fluorescence quantum beats in jet-cooled Anthracene[13], the respective values (in GHz) of $\Delta_{1,2,...4}$ are 3.23, 1.7, 7.57 and 3.7; and $V_{12}$=-0.28, $V_{13}$=-4.24, $V_{14}$=-1.86, $V_{23}$=0.29, $V_{24}$=1.82, $V_{34}$=0.94. When these values are incorporated in Eqn. (2), we obtain the full zero-order Hamiltonian matrix that can simulate the experimental quantum beats (Fig. 3a) upon excitation with a transform-limited Gaussian pulse (i.e., $\dot{\phi}(t) = 0$). Since |0> and |1> do not form a closed two-level system, considerable dephasing occurs during the second half of the Gaussian pulse. Thus, in a coupled multilevel system, simple unchirped pulses cannot be used to generate sequences of $\pi/2$ and $\pi$ pulses, as in NMR. The dark states start contributing to the dressed states, well before the pulse reaches its peak, and results in redistributing the population from the bright state (|1>) into the dark states (Fig. 3a). The situation is worse when we use the tier-model Hamiltonian in Eqn. (3) as the redistribution occurs within the bright states through the participation of the dark states (Fig. 3b).

A linear sweep in frequency of the laser pulse (i.e., $\dot{\phi}(t) = 2b_2t$) can be generated by sweeping from far above resonance to far below resonance (blue to red sweeps), or its opposite. For a sufficiently slow frequency sweep, the irradiated system evolves with the applied sweep and the transitions are "adiabatic". If this adiabatic process is faster than the characteristic relaxation time of the system, such a laser pulse leads to a smooth population inversion, i.e., an adiabatic rapid passage (ARP)[16]. If the frequency sweeps from below resonance to exact resonance with increasing power, and then remains constant, adiabatic *half* passage occurs and photon locking is achieved with no sudden phase shift. However, even under adiabatic *full* passage conditions, Fig.4 shows that there is enough slowing down of the *E* field to result in photon locking over the FWHM of the pulse. These results hold even under certain multiphoton conditions where only an $N^{th}$ ($N \geq 2$) photon transition is possible[11]. Theoretically, scaling the number of dark states is possible as long as there is finite number of states and there are no physical limitations on Stark shifting.

The Quadratic Chirp, i.e., $\dot{\phi}(t) = 3b_3t^2$, is the most efficient in decoupling the bright and dark states as long as the Stark shifting of these states prevail at the peak of the pulse. As the pulse is turned off, the system smoothly returns to its original unperturbed condition (Fig. 5a). This would be a very practical approach of controlling the coupling of the states with realistic pulse shapes. In Fig. 5b, we show that it is possible to initialize the qubits to equal superposition as is required for further quantum operations only with the help of decoherence controlling shaped pulses even when the intramolecular coupling between the states are strong (i.e., for large values of $V_{14}$, $V_{15}$, $V_{24}$, etc.). This is possible since the time dependence can be completely controlled by the experimenter when a shaped pulse is being used.

The cubic term, i.e., $\dot{\phi}(t) = 4b_4t^3$ behaves more like the linear term (Fig. 6). It also decouples the bright and dark states as long as the Stark shifting of these states prevail at the peak of the pulse. However, the oscillatory nature of the "photon-locking" shows that the higher-order terms in the Taylor series involve more rapid changes and fails to achieve perfect adiabatic conditions. As the pulse is turned off, it attempts to invert the bright state population, which quickly dephases, analogous to the linear chirp case. Thus, in an isolated two-level system that does not suffer from the population dephasing, the linear, cubic, and all the higher odd-order terms of the Taylor series (Eqn. (1)) yield inversion of population, while the even-order terms produce self-induced transparency.

For a multilevel system, the induced optical AC Stark-shift by the frequency swept pulse moves the off-resonant coupled levels far from the resonant state leading to an effective decoupling. Under the perfectly adiabatic condition, pulses with the even terms in the Taylor series return the system to its unperturbed condition at the end. In fact, all higher-order odd terms behave in one identical fashion and the even terms behave in another identical fashion. It is only during the pulse, that the Stark-shifting of the dark states are decoupled and IVR restriction is possible in the multi-level situation. In the present calculations, we have used equal values to $b_n$ in Eqn. (1), to bring out the effects of the higher-order terms in the series. In practice, since Eqn.(1) represents a convergent series, only lower-order terms are more important, and since all higher-order terms produce the same qualitative results as the lower-order terms, one needs to consider only up to the quadratic term.

We have already discussed how Hadamard gates can be generated under such adiabatic manipulations. Using the adiabatic coupling schemes as discussed above, it is possible to further construct gates with multilevel systems and such truth tables are shown in Tables I and II. In Table I, we use an ensemble of pseudo-two level system B that can be generated from any IVR multilevel system as discussed here. B can either be in ground (state 0) or excited (state 1) on interacting with control pulse A, which provides robust chirped pulse inversion (condition 1) and the self induced transparency or dark pulse (condition 0). Similarly, in Table II, we consider a three-level system D that can be in ground (state 000) or $1^{st}$ excited (state 010) or $2^{nd}$ excited (state 001) on respective interaction with control pulse C, which provides robust chirped pulse inversion to the $1^{st}$ excited state (condition 010), robust chirped pulse inversion to the $2^{nd}$ excited state (condition 001) and the self induced transparency or dark pulse (condition 000). Such interactions can be considered as pseudo-CNOT gates where the control is in the shaped pulse.

The results are generic and illustrate that adiabatic scheme can be used for control of population transfer for two and three-level systems such that they can result in ensemble gates and for multilevel systems, the intramolecular dephasing can be kept to a minimum for the duration of the "locking" period under adiabatic conditions. In all these cases, since the effect occurs under an adiabatic condition in all these frequency swept pulses, it is insensitive to the inhomogeneity in Rabi frequency. The simulations have been performed with laser pulses with Gaussian, hyperbolic-secant and cosine-squared intensity profiles over a range of intensities. They show identical results of "locking" the population in the chosen excited state of a multilevel system, conforming to the adiabatic arguments that there is hardly any effect of the actual envelope profile. These results are examples of the robustness and utility in the scheme of adiabatic processes that are critical to the adiabatic quantum computing scheme. To our knowledge, these results presented here form the first realistic approaches in the demonstration of the possibility to use ensemble states for developing robust adiabatic quantum computing scheme in multilevel systems.

The author is supported through the Wellcome Trust Senior Research Fellow program of the Wellcome Trust Foundation (UK) and the Swarnajayanti Fellow scheme under the Dept. of Science and Technology, Govt. of India. He also wishes to thank the Ministry of Information Technology, Govt. of India, for partial funding of the research results presented here.

## Figures

**FIG. 1.** (a) Schematic of a three-level system with two possible transitions modeling atomic sodium and rubidium atoms. A short pulse has enough bandwidth ($\Delta\omega$) to excite both the possible states. (b) A transform-limited 2 ps Gaussian pulse having enough bandwidth ($\Delta\omega$) interacts with a model three level atomic system in a single photon mode or in a multiphoton condition and the population evolution shows that no selectivity in population transfer is possible. (c) A linearly swept Gaussian pulse can generate selective inversion and depending on the sign of the frequency chirp (whether (i) red to blue or (ii) blue to red) can selectively invert the population under the adiabatic limit.

**FIG. 2.** (a) Schematic of IVR for Anthracene molecule from Ref. [9] based on data extracted from experimental measurements in Ref. [13]. (b) Model tier level coupled IVR system common in many polyatomic systems. This is also a common coupled level system for rovibrational states.

**FIG. 3.** (a) A transform-limited Gaussian pulse interacts with a model Anthracene molecule in a single photon mode or in a multiphoton condition. (b) Due to strong coupling of the excited states, 1>, |2> and |3> to |4>, |5>, etc., population of the excited states are highly modulated even during the period of a simple Gaussian excitation pulse for the tier-model system given in Fig.2b.

**FIG. 4.** A linearly swept Gaussian pulse can generate "photon-locking". The evolution of the dressed state character is unchanged while locking occurs but as the pulse is turned off, the eigen-energy curves cross and the bright state population quickly dephases.

**FIG. 5.** (a) The Quadratic Chirped Gaussian pulse is the most efficient in decoupling the bright and dark states during the pulse. The eigen-energy curves and the corresponding evolution of the dressed state character shows that the entire process is highly adiabatic. (b) The Quadratic Chirped Gaussian pulse effectively generates equal superposition of the excited states (|1>, |2>, etc.) during the period of the shaped pulse for the tier-model system given in Fig.2b.

**FIG. 6.** Effect of a Cubic Chirped Gaussian pulse is similar to the linearly swept pulse (Fig.3), although the evidence of population oscillation indicates that this chirp is not as adiabatic as the linear chirp. The eigen-energy curves cross towards the end of the pulse and the bright state population gets redistributed.

## Table I

Adiabatic Gates with Chirped Pulses for pseudo two-level system

## Table II

Adiabatic Gates with Chirped Pulses for three-level system

Table I

| Shaped pulse | A | B | A⊕B |
|---|---|---|---|
| "Inverting" pulse | 1 | 1 | 0 |
| | 1 | 0 | 1 |
| "Dark" pulse | 0 | 1 | 1 |
| | 0 | 0 | 0 |

Table II

| Shaped pulse | C | D | C⊕D |
|---|---|---|---|
| "Inverting" pulse selective to 1st excited | 010 | 100 | 010 |
| | 010 | 010 | 100 |
| | 010 | 001 | 001 |
| "Inverting" pulse selective to 2nd excited | 001 | 100 | 001 |
| | 001 | 010 | 010 |
| | 001 | 001 | 100 |
| "Dark" pulse | 000 | 100 | 100 |
| | 000 | 010 | 010 |
| | 000 | 001 | 001 |

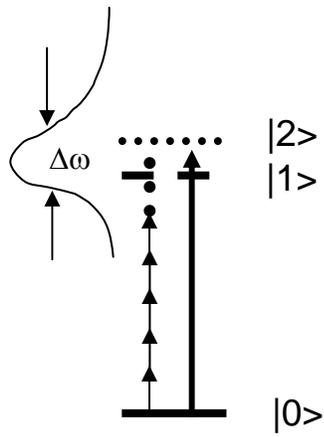

Figure 1 (a)

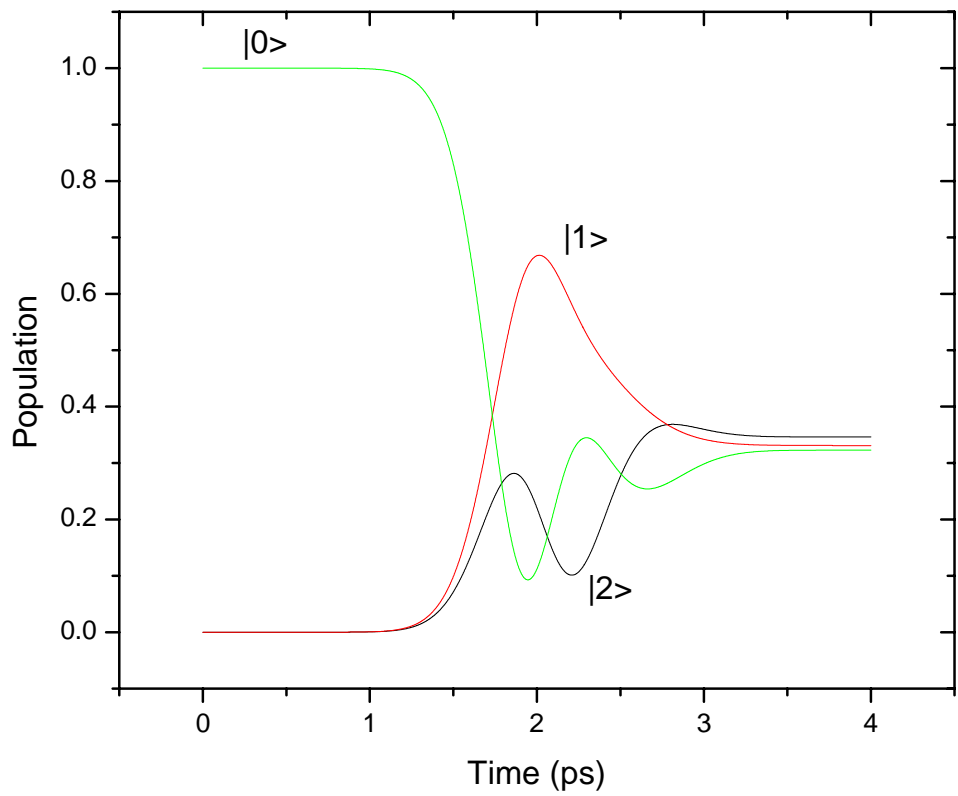

Figure 1 (b)

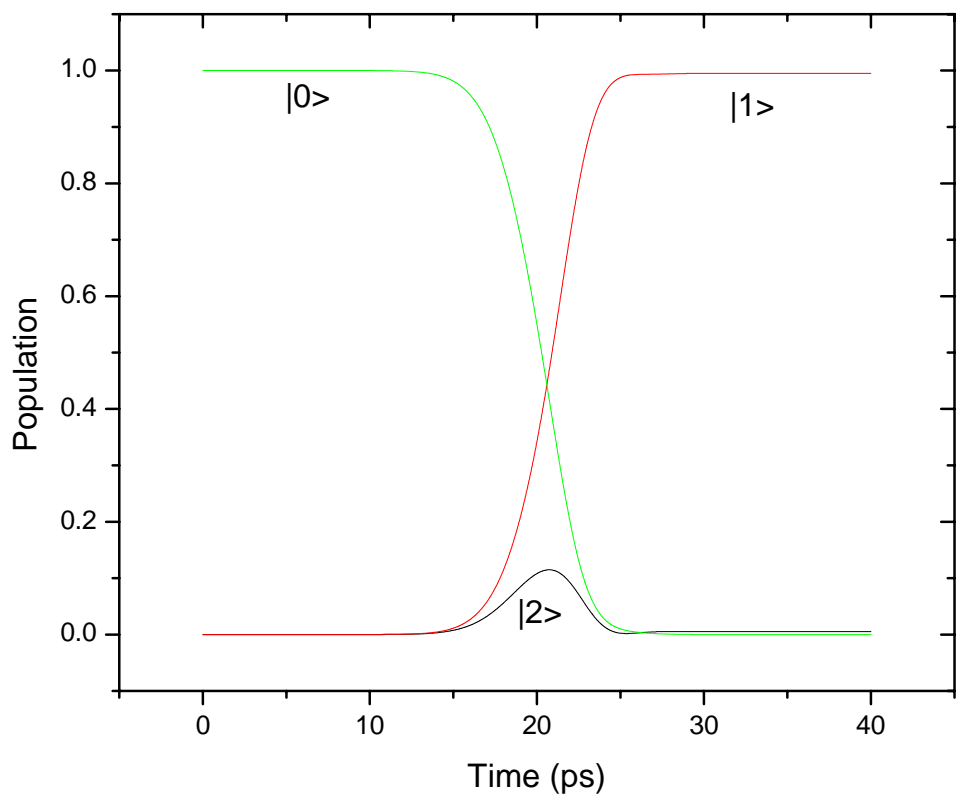

(i)

Figure 1 (c)

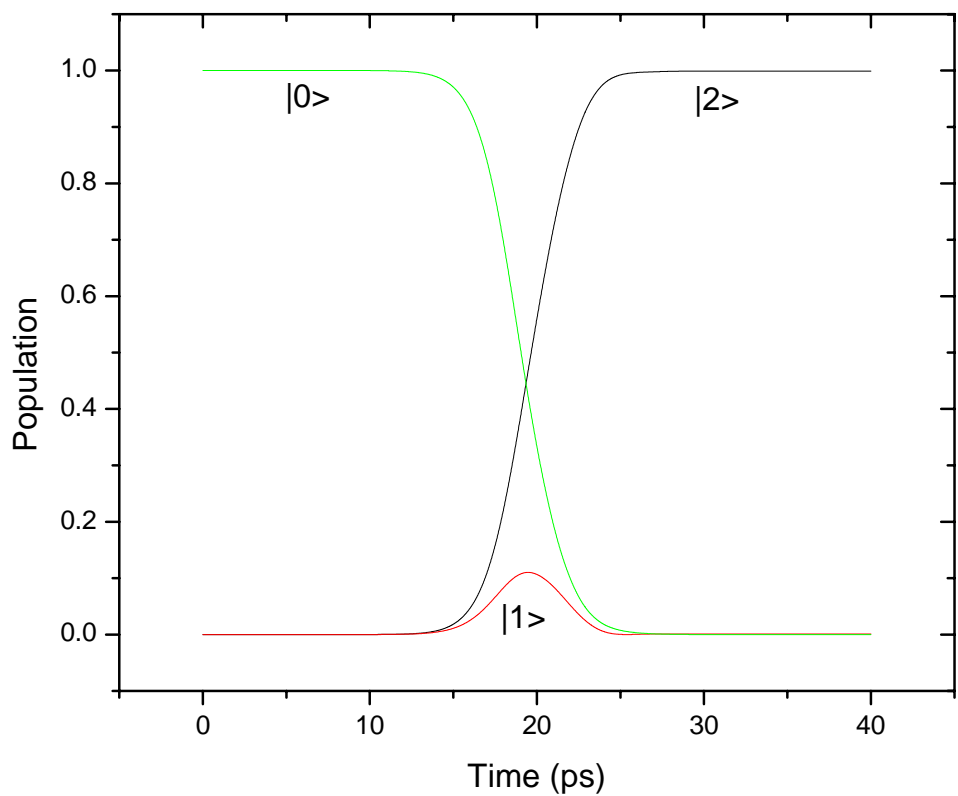

(ii)

Figure 1 (c)

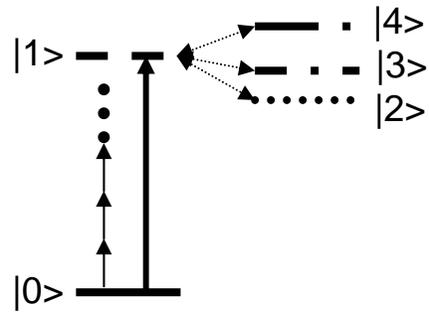

Figure 2 (a)

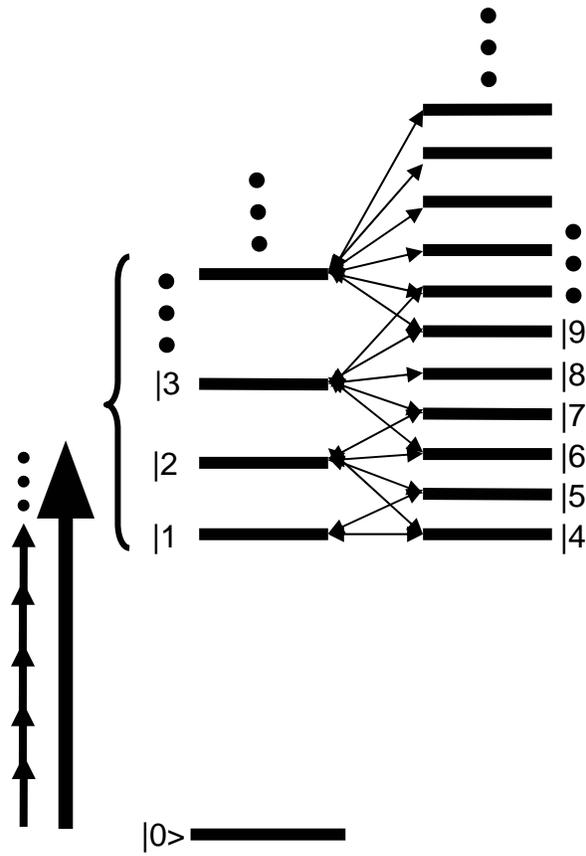

Figure 2 (b)

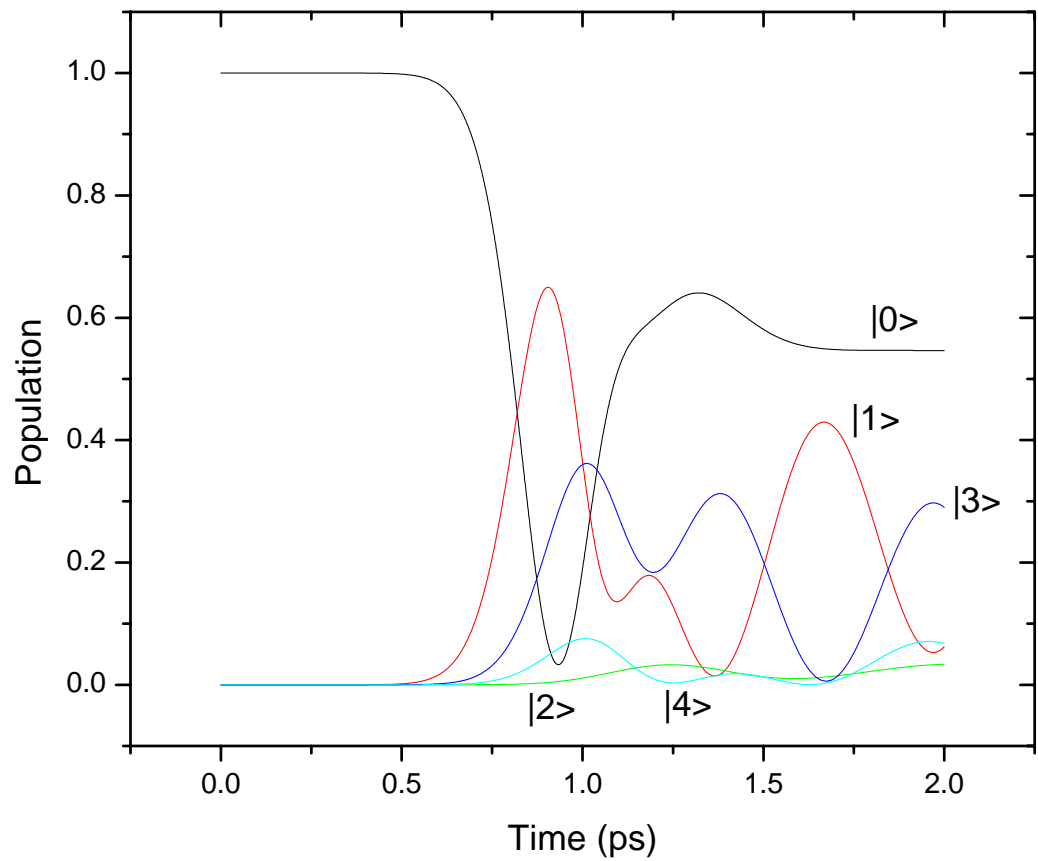

Figure 3 (a)

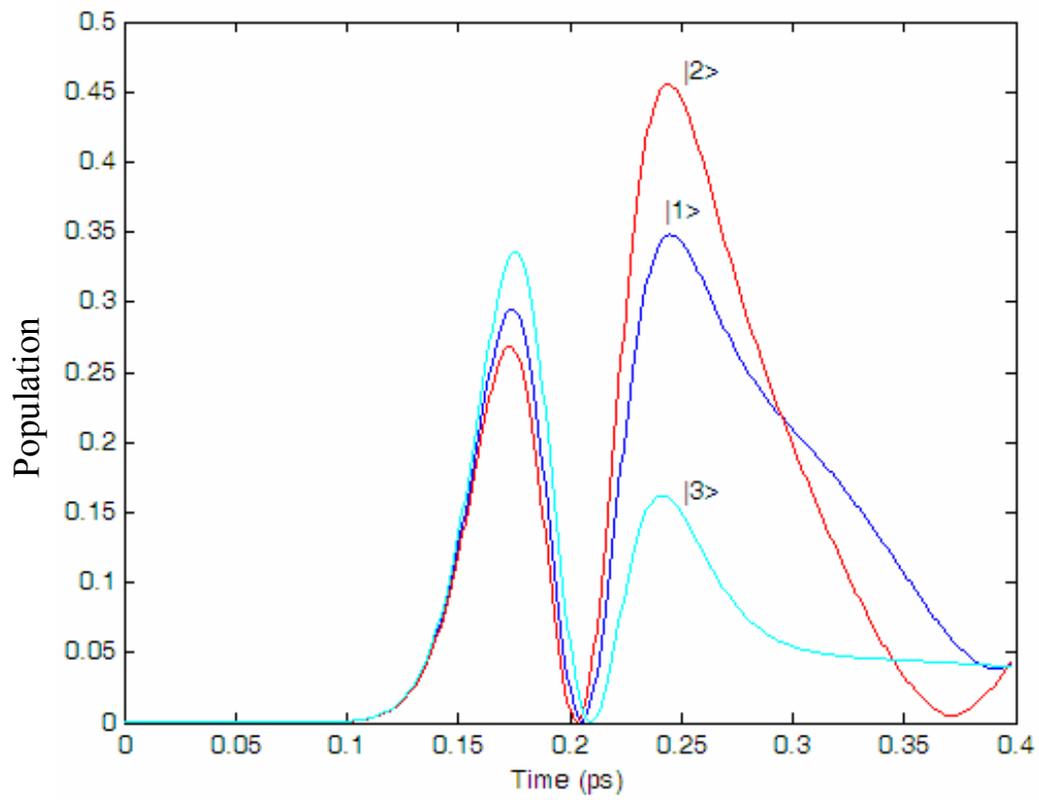

Figure 3 (b)

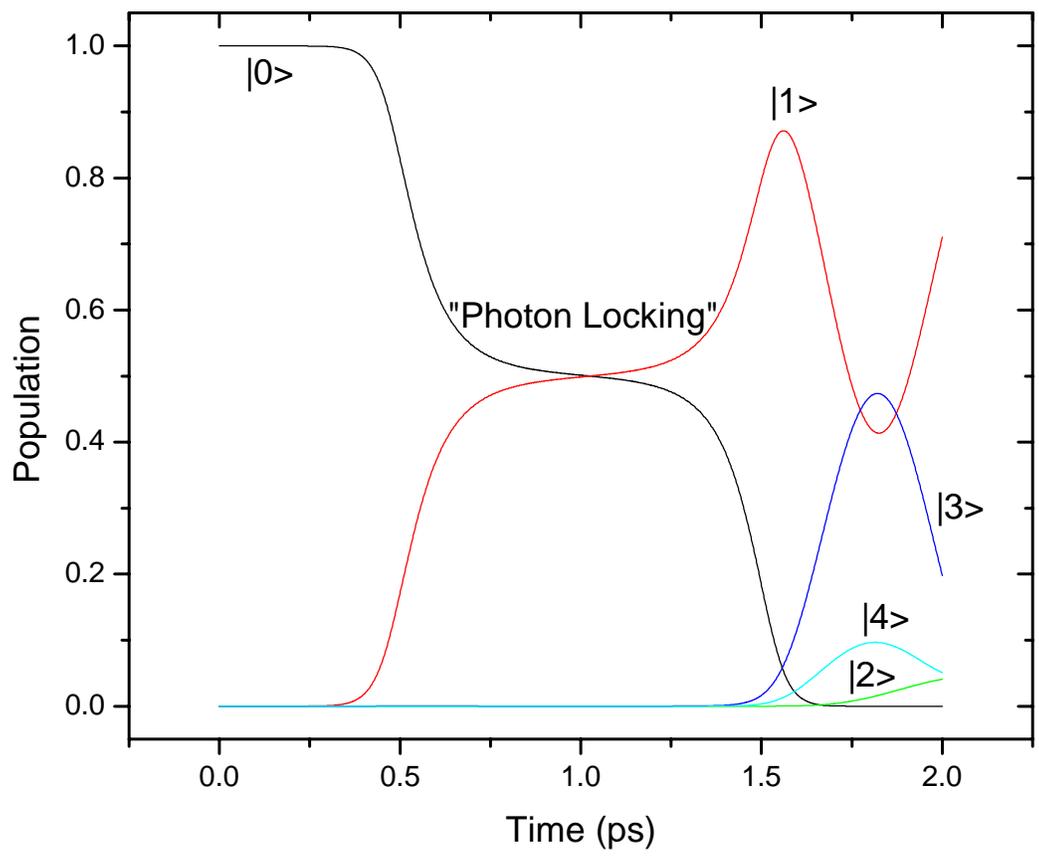

Figure 4

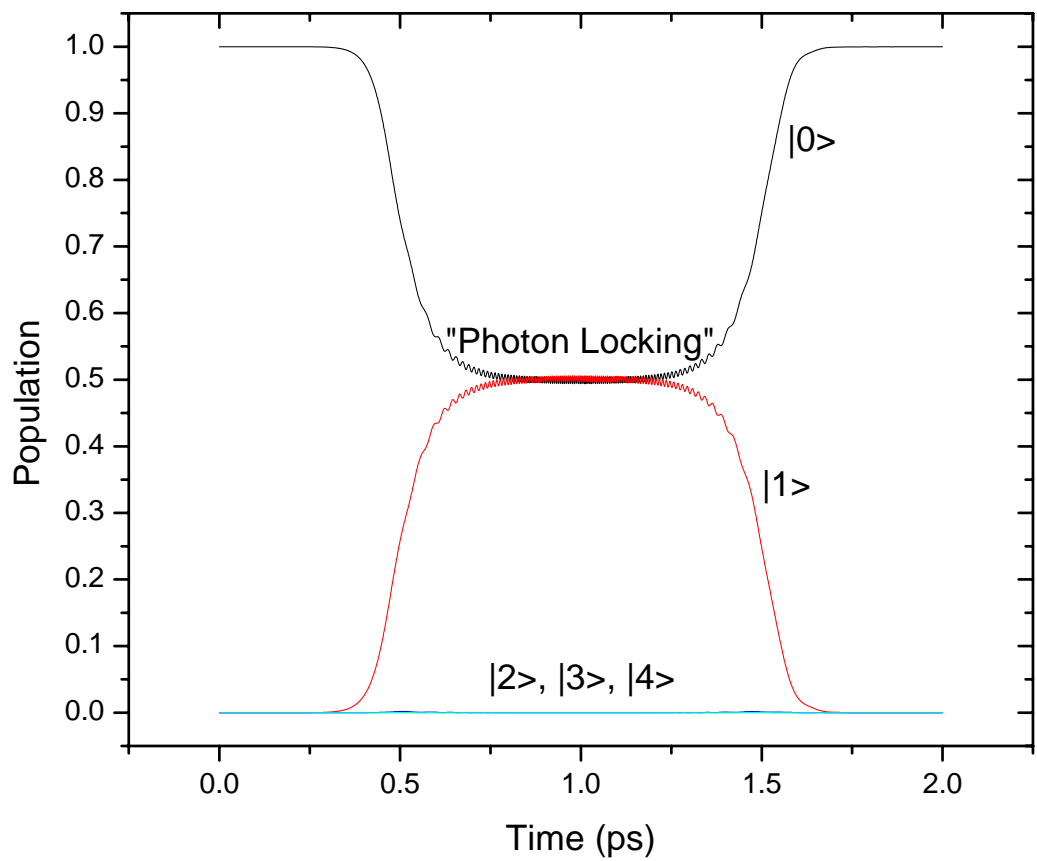

Figure 5 (a)

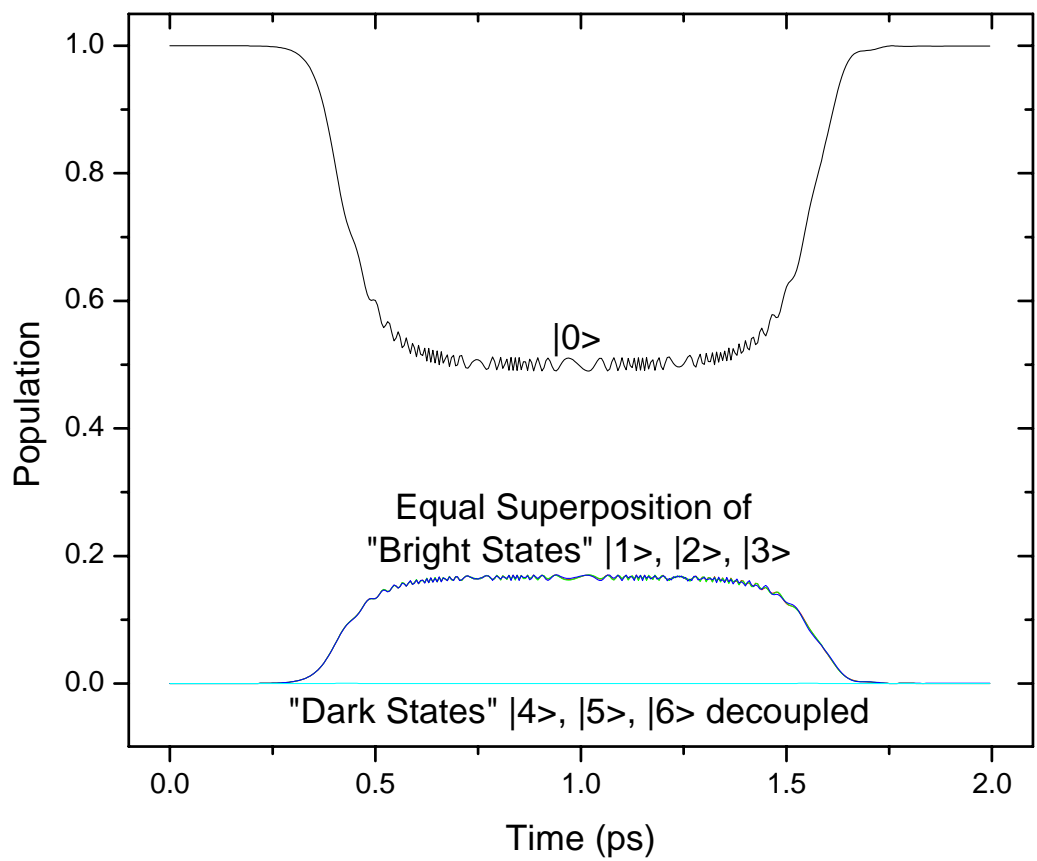

Figure 5 (b)

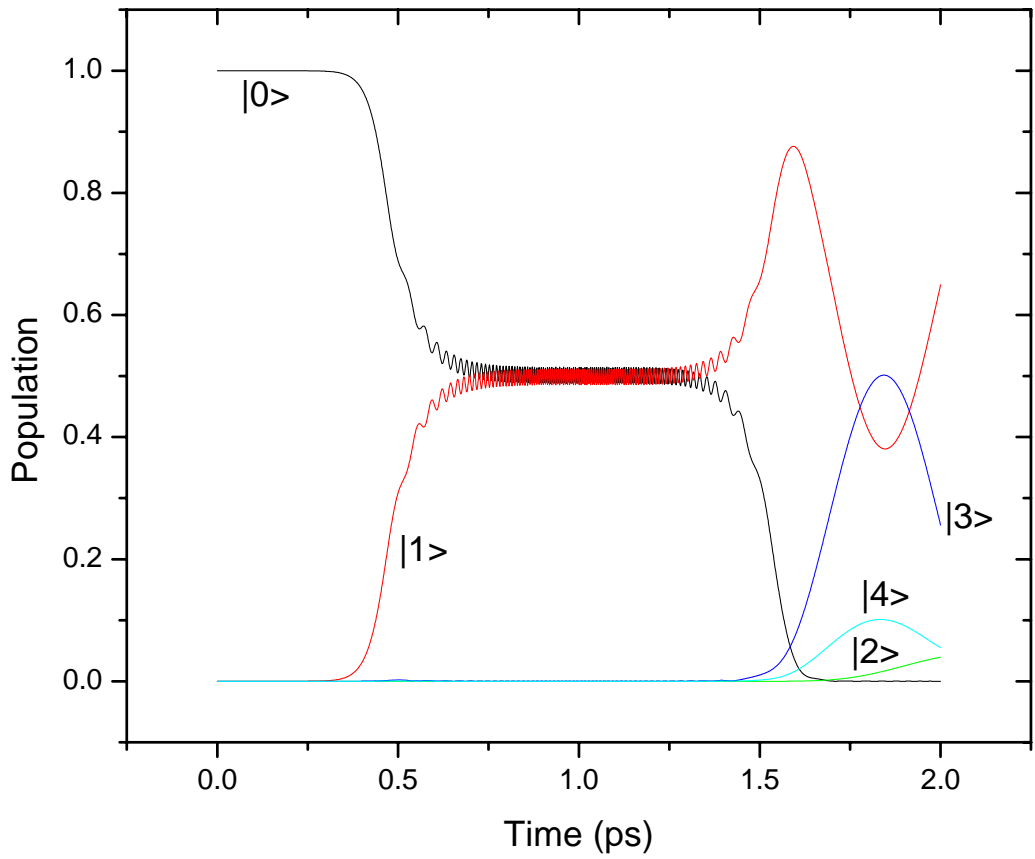

Figure 6